\UseRawInputEncoding
\documentclass[pra,twocolumn,showpacs,floatfix]{revtex4}
\usepackage{graphicx}
\usepackage{color}
\usepackage{amsmath}
\begin{document}

\title{Stationary states of Bose-Einstein condensed atoms rotating in an asymmetric ring potential}

\author{M. \"{O}gren$^{1,2}$, Giannis Drougakis$^{3,4}$, Giorgos Vasilakis$^3$, Wolf von Klitzing$^{3}$, 
and G. M. Kavoulakis$^{2,5}$}

\affiliation{$^{1}$School of Science and Technology, \"{O}rebro University, 70182 \"{O}rebro, Sweden \\
$^2$Hellenic Mediterranean University, P.O. Box 1939, GR-71004, Heraklion, Greece \\
$^3$Institute of Electronic Structure and Laser, Foundation for Research and Technology-Hellas, Heraklion 70013,
Greece \\
$^4$Department of Materials Science and Technology, University of Crete, Heraklion 70013, Greece \\
$^5$HMU Research Center, Institute of Emerging Technologies, P.O. Box 1939, GR-71004, Heraklion, Greece}

\date{\today{}}

\begin{abstract}

We consider a Bose-Einstein condensate, which is confined in a very tight toroidal/annular trap, in
the presence of a potential, which breaks the axial symmetry of the Hamiltonian. We investigate the 
stationary states of the condensate, when its density distribution co-rotates with the symmetry-breaking
potential. As the strength of the potential increases, we have a gradual transition from vortex excitation 
to solid-body-like motion. Of particular importance are states where the system is static and yet it has 
a nonzero current/circulation, which is a realization of persistent currents/reflectionless potentials. 

\end{abstract}
\pacs{05.30.Jp, 03.75.−b, 03.75.Kk} \maketitle

\section{Introduction}

One of the many applications of the field of cold atomic gases is
that of ``atomtronics" \cite{roadmap,roadmap2}. With this term we 
mean the use of guided cold atoms, in devices which resemble those 
of electronic systems, with potential applications. In such an effort, 
it is crucial to be able to manipulate and guide the atoms. Actually, 
numerous experimental groups have managed to build topologically-nontrivial 
potentials, i.e., annular, or toroidal traps \cite{Sauer,Kurn,Arnold,Olson,
Phillips1,Heathcote,Henderson,Foot,GK,Moulder,Zoran,Ryu,WVK,hysteresis,hyst2,
Perin,WVK2}, in an effort to realize such devices. Clearly, setting the atoms 
in motion and guiding them is essential. Numerous experiments have been 
performed in this direction (and probably they are too many to cite.) Here 
we just mention a few experiments in annular and/or toroidal potentials, 
which have managed to create persistent currents, to observe hysteresis, 
to observe supersonic velocities, etc. \cite{Kurn,Olson,Phillips1,Foot,GK,
Moulder,Zoran,Ryu,WVK,hysteresis,hyst2,Perin,WVK2}.

The study of the flow of atoms in the presence of an external potential,
which breaks the rotational invariance -- in the case of an
annulus/torus, or the translational invariance -- in the case of
a waveguide, is of fundamental importance. The breaking of 
rotational/translational symmetry is always present in any real system. 
This is either due to the unavoidable existence of ``weak" irregularities, 
or due to potentials which we create on purpose, in order to guide and 
manipulate the atoms. Therefore, in the study of this problem, it is 
necessary to consider the effect of an external potential, which is not 
``weak", in general.

Motivated by the above remarks, we study the rotational properties of 
a Bose-Einstein condensed gas of atoms in the presence of an external
potential \cite{theory1,theory2,theory3,theory33,theory4,theory45,theory44,expr,theory55,theory5}. 
We assume for simplicity that the atoms are confined in a ring potential, 
having in mind a very narrow annulus, or torus, where the transverse 
degrees of freedom are frozen. For a typical value of the
chemical potential $\mu/\hbar = 10$ kHz, the frequency in the transverse
direction has to be larger/much larger that this, in order to achieve
the conditions of quasi-one-dimensional motion. We stress that such
conditions have been realized already roughly 20 years ago, in the 
experiment of Ref.\,\cite{kett1d}, in an elongated trap. 

An interesting aspect of this problem is the combined effect of the 
periodic boundary conditions that we impose in our solutions, with the 
potential, which breaks the axial symmetry of the Hamiltonian. We look 
for stationary solutions in the rotating frame, assuming that the density 
wave associated with the rotating atoms has the same angular velocity as 
the symmetry-breaking potential. Interestingly enough, as we demonstrate 
below, even a very weak potential affects the rotational response of the 
system drastically, as compared to the axially-symmetric case.

As mentioned above, the problem that we have studied has 
been investigated thoroughly by several authors. The novelty of our 
results relies, first of all, on the fact that we work with a fixed angular 
momentum. This allows us to get a very clear picture of the change in the
rotational behaviour of this driven system as the strength of the driving
potential is varied. In addition, we derive limiting analytical results,
which not only agree with the numerical results, but also provide insight 
into this problem. 

To state just a few of our most important results, first of all, we have 
derived an expression for the moment of inertia of the system for a 
``weak" external potential (compared with the chemical potential), for 
small enough values of the angular momentum, where the system behaves 
as a solid body. With increasing angular momentum though, the system 
turns to superfluid rotational behaviour. When the external potential 
becomes sufficiently strong, the system turns to solid-body-like motion 
for all values of the angular momentum. Finally, we have identified 
states which correspond to a static external potential. Obviously, 
the density distribution of the cloud is inhomogeneous in this case 
and furthermore this is not a driven system. Still, these states 
correspond to (metastable, non-decaying) persistent currents, or, in 
an alternative terminology, we have the realization of reflectionless 
potentials.

In what follows below we first present our model in Sec.\,II, 
which is based on the mean-field approximation. In Sec. III we describe 
Bloch's theorem. In Secs. IV and V we evaluate analytically and numerically 
the dispersion relation in the absence and in the presence of an external 
potential. In Sec. VI we examine the effect of the potential on the observables. 
In Sec.\,VII we connect our results with some relevant experimental 
values. Finally, in Sec. VIII we present a summary of our results and the main 
conclusions of our study. 

\section{Model}

Let us assume that we have Bose-Einstein condensed atoms, which are confined 
in a ring potential, with a radius $R$. Within the mean-field approximation, the 
order parameter of the condensate $\Phi(z,t)$ then satisfies the equation
\begin{eqnarray}
  i \hbar \frac {\partial \Phi} {\partial t} =
- \frac {\hbar^2} {2 M} \frac {\partial^2 \Phi} {\partial z^2}
+ V(z, t) \Phi + g N |\Phi|^2 \Phi,
\label{gpe}
\end{eqnarray}
with $\int |\Phi|^2 \, dz = 1$. Here, $z$ is the spatial coordinate, with $-\pi R \le z
\le \pi R$, $M$ is the atom mass, $V(z,t)$ is a potential which acts along the ring, $N$ 
is the atom number, and $g$ is the matrix element for elastic atom-atom collisions. We 
assume that $g>0$, i.e., repulsive interatomic interactions. 

We look for travelling-wave solutions, under periodic boundary conditions, with a velocity 
of propagation $u$, i.e., $\Phi(z,t) = \Psi(x) e^{-i \mu t/\hbar}$, where $x=z-ut$ and $\mu$ 
is the chemical potential. Furthermore, in order to get a time-independent equation, we assume 
that the potential moves with the same velocity $u$ as the wave. Under these two conditions 
Eq.\,(\ref{gpe}) becomes,
\begin{eqnarray}
  - i \hbar u \frac {d \Psi} {d x} =
- \frac {\hbar^2} {2 M} \frac {d^2 \Psi} {d x^2}
+ [V(x) + g N |\Psi|^2 - \mu] \Psi.
\label{gpesss}
\end{eqnarray}
The above equation results from the minimization of the following extended energy functional
\begin{eqnarray}
  {\cal E}(\Psi, \Psi^*) = 
  - \frac {\hbar^2} {2 M} \int \Psi^* \frac {d^2 \Psi} {d x^2} d x
+ \int |\Psi|^2 V(x) \, d x 
\nonumber \\
+ \frac 1 2 g N \int |\Psi|^4 \, d x  
- \mu \int |\Psi|^2 \, d x 
 - \Omega (- i) \hbar R \int \Psi^* \frac {d \Psi} {d x} \, d x,
\nonumber \\
\label{gpessss}
\end{eqnarray}
where $\Omega = u/R$ is the angular velocity of the density wave, and of the external potential. 
 
One may think that because of the potential $V(x)$ the angular momentum is not conserved. In the 
present problem, though, we make the rather drastic assumption that the potential moves with the 
solitary wave and we derive travelling-wave solutions. This implies that -- effectively -- we still 
have axial symmetry, since there is no preferable point along the ring, with the position of the 
center of the solitary wave and of the extremum of the potential being arbitrary. As a result, the
angular momentum is conserved (and Bloch's theorem is valid, as analysed in the following section)
\cite{remark}. 

Equation (\ref{gpessss}) thus corresponds to the minimization of the energy under a fixed expectation 
value of the angular momentum and a fixed atom number, with $\mu$ and $\Omega$ being the corresponding 
Lagrange multipliers. An immediate consequence of Eq.\,(\ref{gpessss}) is also that $\Omega = dE(\ell)/d 
(\ell \hbar)$, where $E(\ell)$ is the dispersion relation and $\ell \hbar$ is the angular momentum per 
particle. 

As a result, while $\ell$ may take any value, $\Omega$ is determined by the problem. To get some insight, 
we stress that the angular momentum is associated with the superfluid velocity, i.e., with the gradient 
of the phase of the order parameter. The phase, in turn, has the only restriction that its difference 
between $x = -\pi R$ and $x = \pi R$ has to be an integer multiple of $2 \pi$, due to the periodic 
boundary conditions. On the other hand, as mentioned above, $\Omega$ is determined by the dispersion 
relation. As we show below, unless the potential is strong enough, in which case we have solid-body-like 
motion, the dispersion relation has a quasi-periodic behaviour. As a result, $\Omega$ shows a 
quasi-periodicity, too, which gives rise to the bounds mentioned above. 

\section{Bloch's theorem}

As it was first pointed out by Bloch \cite{FB}, when we have rotational invariance for some fixed 
value of the angular momentum $\ell \hbar$, the dispersion relation may be written in the form
\begin{eqnarray}
  E(\ell) = \epsilon \ell^2 + e(\ell),
\label{drel}
\end{eqnarray}
where $\epsilon = \hbar^2/(2 M R^2)$ is the usual kinetic energy associated with the motion of the 
atoms around the ring. Also, $e(\ell)$ is periodic, with a period equal to unity, and is symmetric 
around $\ell = 1/2$, i.e., $e(\ell) = e(1 - \ell)$, for $0 \le \ell < 1$. 

Denoting as $e'(\ell)$ the derivative of $e(\ell)$ with respect to $\ell \hbar$, since $e'(\ell) = 
- e'(1-\ell)$, therefore $e'(\ell = 1/2)= 0$. As a result, we conclude for $\Omega$ that
\begin{eqnarray}
 \Omega(\ell) = \frac {u(\ell)} R = \frac 1 {\hbar} \frac {d E(\ell)} {d \ell} = \frac {2 \epsilon} {\hbar} \ell 
 + e'(\ell).
 \label{resom}
\end{eqnarray}
Clearly $\Omega(\ell) + \Omega(1 - \ell) = 2 \epsilon/\hbar$ and as a result $\Omega(\ell = q/2) = 
q \epsilon/{\hbar}$, where $q = 1, 3, 5, \dots$. As we argued earlier, Bloch's theorem is still valid in the 
present problem. 

\section{Dispersion relation in the absence of any potential}

In the absence of an external potential, for integer values of the angular momentum, i.e., $\ell \hbar = q 
\hbar$, with $q$ an integer, the lowest-energy state of the system has a homogeneous density distribution 
and is simply in the state $\phi_q(x) = e^{i q x/R}/\sqrt{2 \pi R}$. 

Let us now examine the behaviour of the order parameter around $\ell = q$, focusing on the limit $\ell 
\to q^+$, with our small parameter being $\ell - q$. In this limit the order parameter is, to leading order 
in $\ell - q$, with the amplitudes $c_m$ assumed to be real,
\begin{eqnarray}
  \Psi_q = c_{q-1} \phi_{q-1} + c_q \phi_q + c_{q+1} \phi_{q+1}.
\end{eqnarray}
We have to impose the usual two constraints on $\Psi_q$, i.e., the normalization, $c_{q-1}^2 + c_q^2 + 
c_{q+1}^2 = 1$ and also the constraint of a fixed angular momentum, $c_{q+1}^2 - c_{q-1}^2 = \ell - q$. 
It is convenient to parametrize $c_{q-1}$ and $c_{q+1}$ as $c_{q-1} = \sqrt{\ell-q} \sinh \theta$, and
$c_{q+1} = \sqrt{\ell-q} \cosh \theta$. The energy of the system is then given by
\begin{eqnarray}
 \frac {E_q} {N \epsilon} = c_{q-1}^2 + c_{q+1}^2 + \frac {\delta} 2 (c_{q-1}^4 + c_q^4 + c_{q+1}^4 + 4 c_{q-1}^2 
 c_q^2 +
 \nonumber \\
 + 4 c_{q-1}^2 c_{q+1}^2 + 4 c_{q}^2 c_{q+1}^2 - 4 c_q^2 |c_{q-1}| |c_{q+1}|),
 \label{varen}
\end{eqnarray} 
where $\delta/2$ is the ratio between the interaction energy of the homogeneous state, $g n_0/2$, with 
$n_0 = N/(2 \pi R)$ and $\epsilon$. Keeping the leading-order terms, 
\begin{eqnarray}
   \frac {E_q} {N \epsilon} - \frac {\delta} 2 
   \approx q (1-q) + (2 q - 1) \ell + 2 c_{q+1}^2 + \delta (c_{q+1} - c_{q-1})^2.
   \nonumber \\
\end{eqnarray}
Minimizing this expression, it turns out that 
\begin{eqnarray}
 \frac {E_q} {N \epsilon} - \frac {\delta} 2 = q (2 \ell - q) + \sqrt{2 \delta + 1} \, (\ell-q).
 \label{disprell}
\end{eqnarray} 
Furthermore, $c_{q-1}$ and $c_{q+1}$ have opposite signs and both scale as $\sqrt{\ell-q}$, i.e.,
\begin{eqnarray}
  |c_{q \pm 1}| = \frac 1 2 {\sqrt{\ell - q}} \frac {\sqrt{2 \delta + 1} \pm 1} {(2 \delta + 1)^{1/4}}.
\label{cs}
\end{eqnarray}
One of the important conclusions of this analysis is that the dispersion relation is linear in $\ell-q$, 
with a slope equal to $(2q + \sqrt{2 \delta + 1}) \epsilon/\hbar$ for $\ell \to q^+$, as seen in the top
right plot of Fig.\, (first, third, and fifth from the left, green, dashed lines). For 
$\ell \to q^-$, the slope is equal to $(2q - \sqrt{2 \delta + 1}) \epsilon/\hbar$ (second and 
fourth from the left, green, dashed lines), and as a result there is a discontinuity in the slope, i.e., 
in $\Omega$, which is equal to $2 \sqrt{2 \delta + 1} \epsilon/\hbar$, as seen in Figs.\,3 and 4 
(for $V(x) = 0$). 

In addition to the analytic results which are presented above, we have performed numerical simulations
of this problem, minimizing the extended energy functional ${\cal E}$ of Eq.\,(\ref{gpessss}) \cite{PS}, 
both in the absence and in the presence of an external potential, of the form
\begin{eqnarray}
  V(x) = V_0 e^{-x^2/(2 w_0^2)}.
  \label{expot}
\end{eqnarray} 
Figure 1 shows the result of this calculation (for $V_0 = 0$, and also $V_0 \neq 0$, which is examined below), 
for some representative values of the angular momentum $\ell \hbar$. Also, $\delta$ is chosen to
be equal to 7.5 both in this figure, as well as in all the other ones. When there is no external potential and 
$\ell$ is an integer, then the density is homogeneous. In the interval $q \le \ell \le q+1/2$ (with $q$ an integer) 
as $\ell$ increases, the minimum of the density decreases, dropping down to zero for $\ell = q+1/2$, when we get the 
``dark" solitary waves. We stress that, in a finite ring, a dark wave is not static, but rather it has a finite 
velocity of propagation \cite{comm}, as pointed out in Ref.\,\cite{aaa}. Then, for $q+1/2 \le \ell \le q+1$, the 
minimum of the density increases with increasing $\ell$.

Turning to the phase of the order parameter, this satisfies the obvious periodic boundary conditions. It also 
develops a jump, for exactly $\ell = q+1/2$, i.e., in the case of the dark solitary waves. Finally, the dispersion 
relation always has a negative curvature, with discontinuities in its first derivatives -- i.e., in $\Omega$ -- at 
the integer values of $\ell=q$, in full agreement with the analytic results presented above. 

\begin{figure}
\includegraphics[width=4cm,height=4.cm]{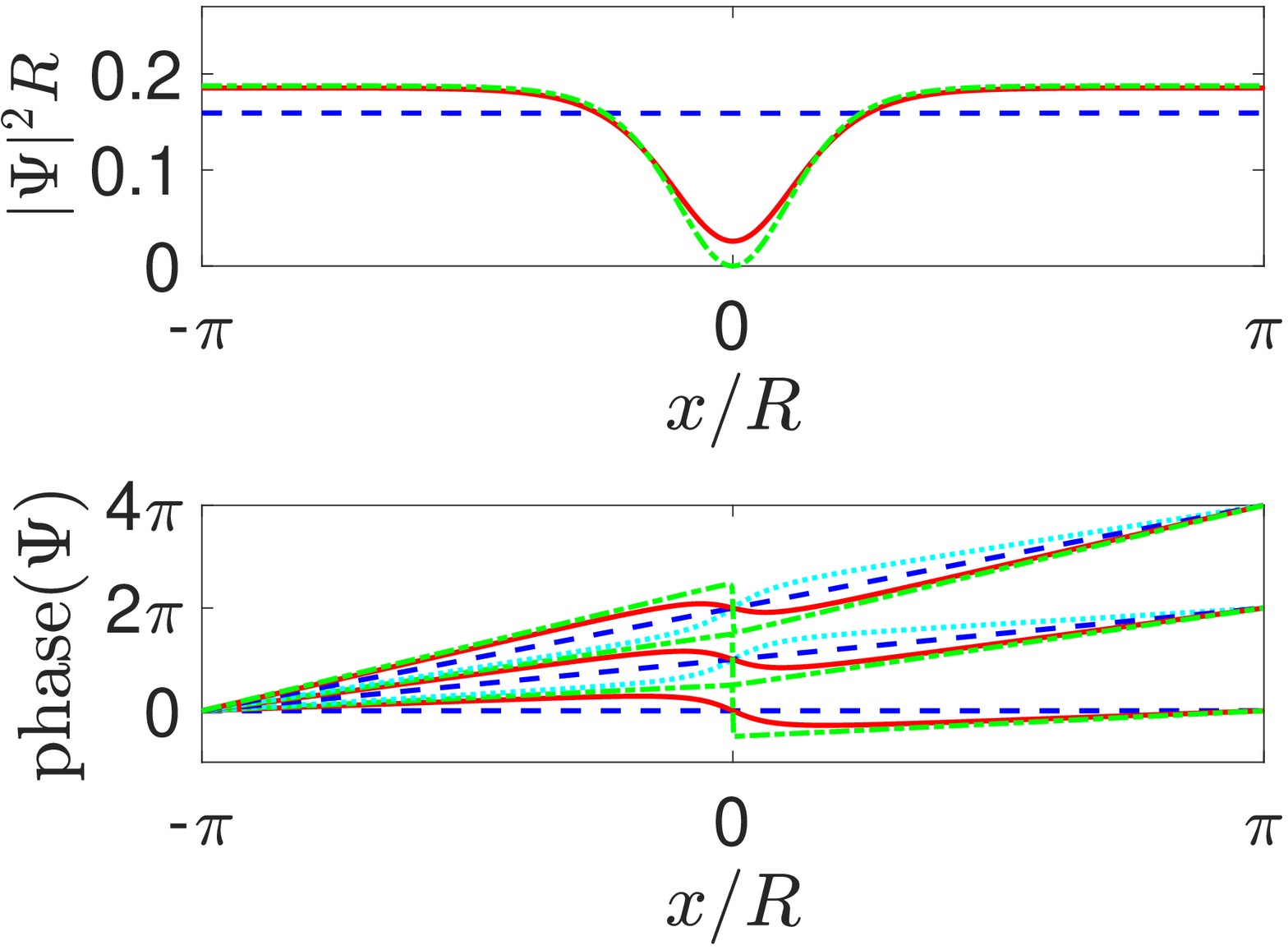}
\includegraphics[width=4cm,height=4.cm]{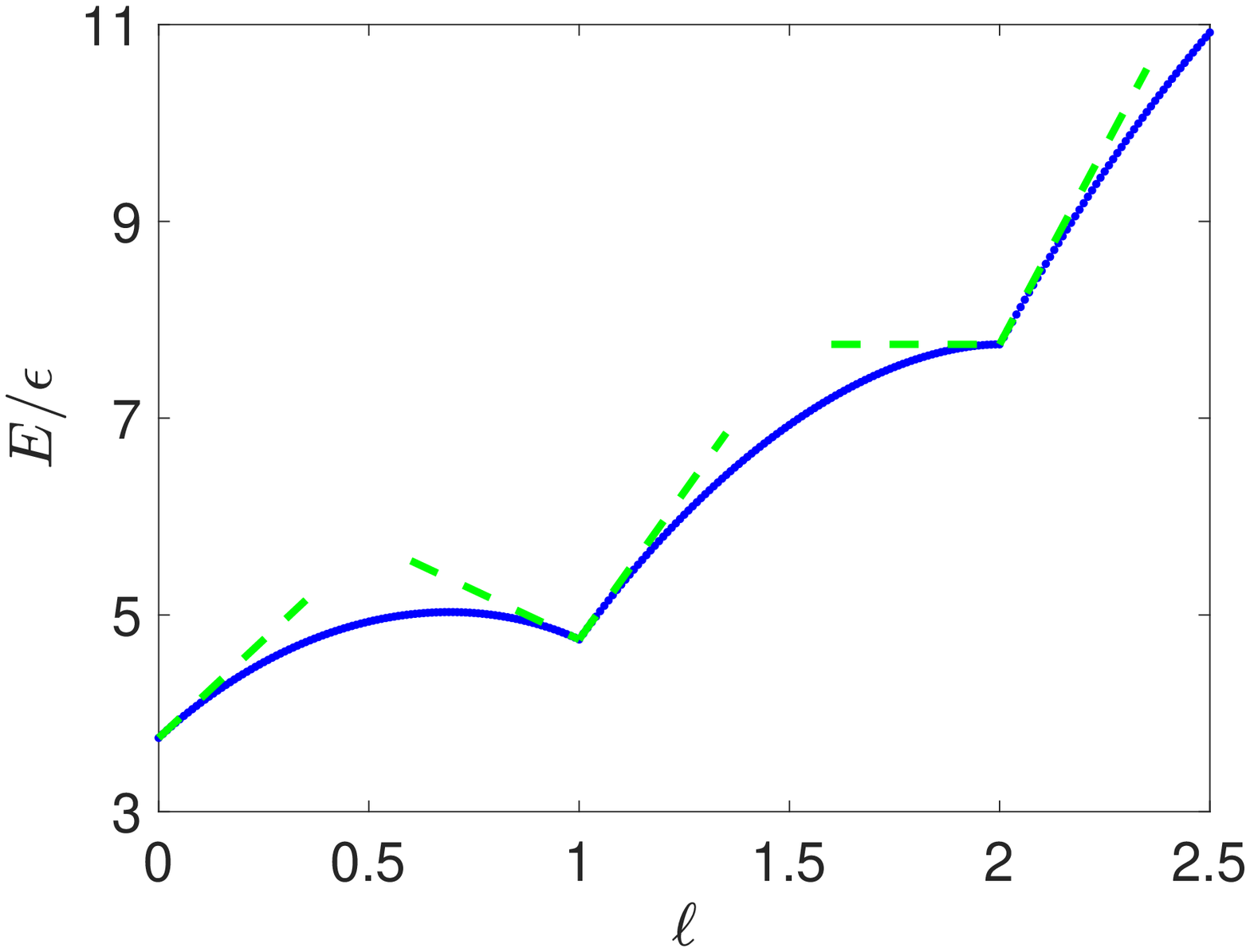}
\includegraphics[width=4cm,height=4.cm]{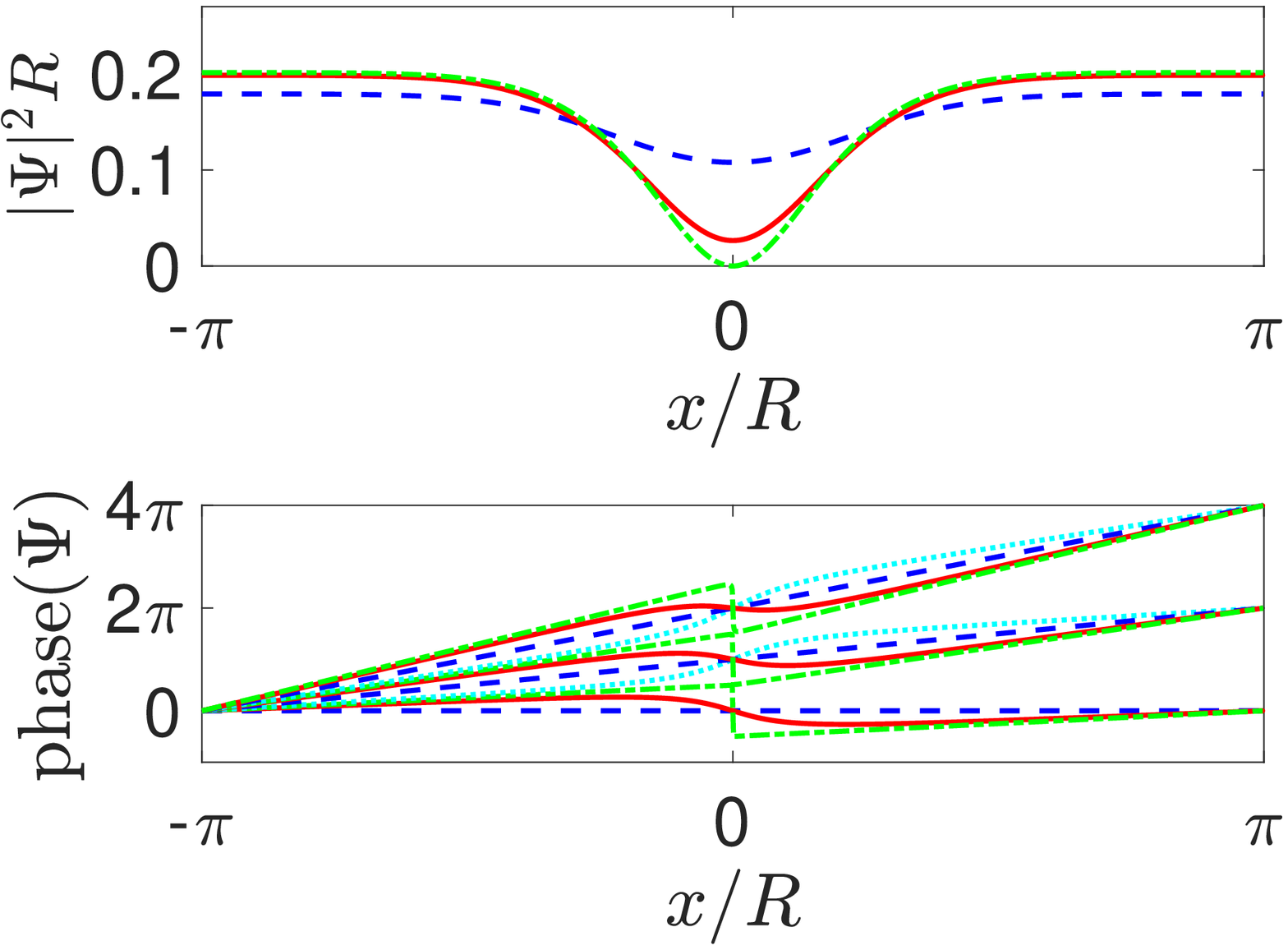}
\includegraphics[width=4cm,height=4.cm]{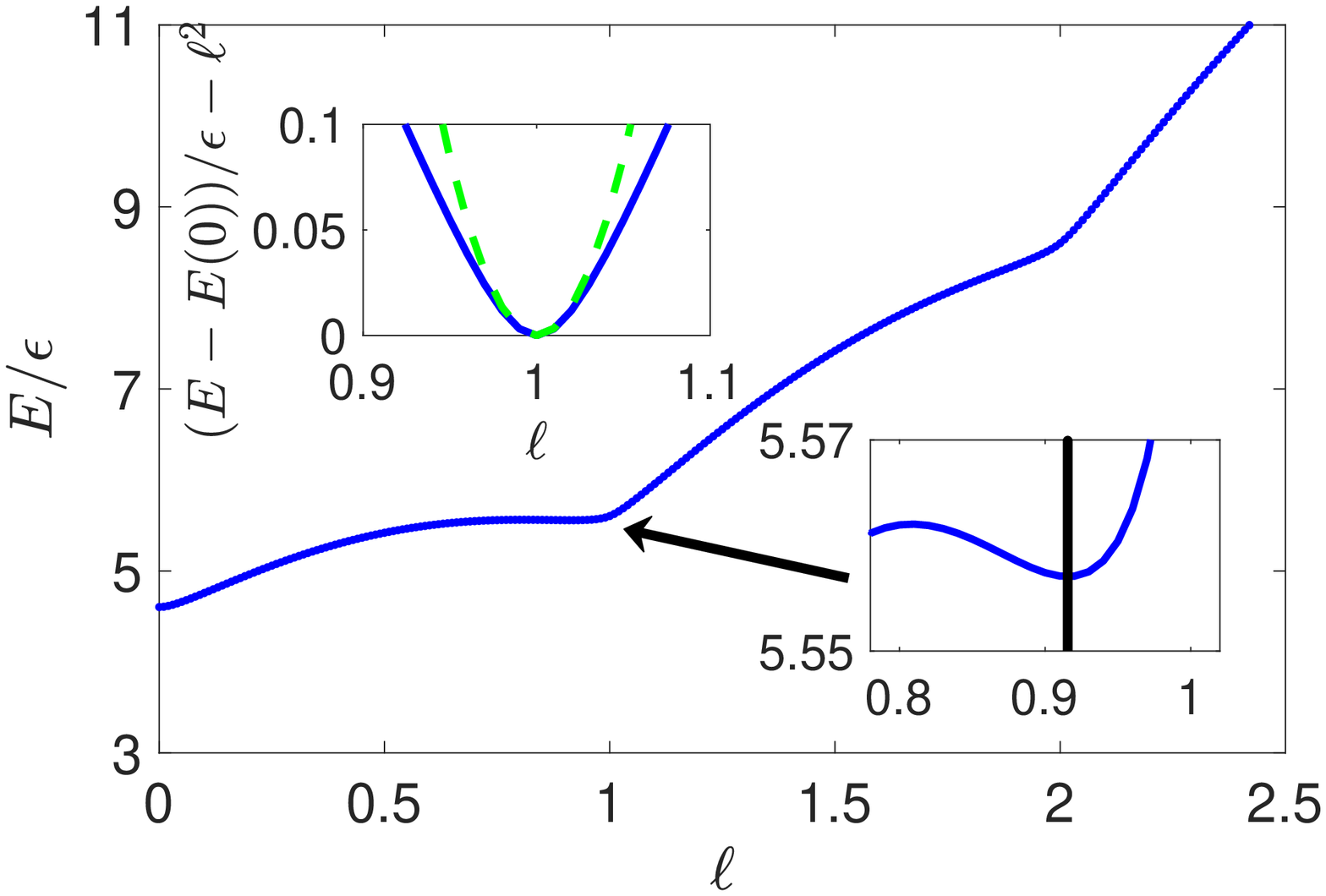}
\includegraphics[width=4cm,height=4.cm]{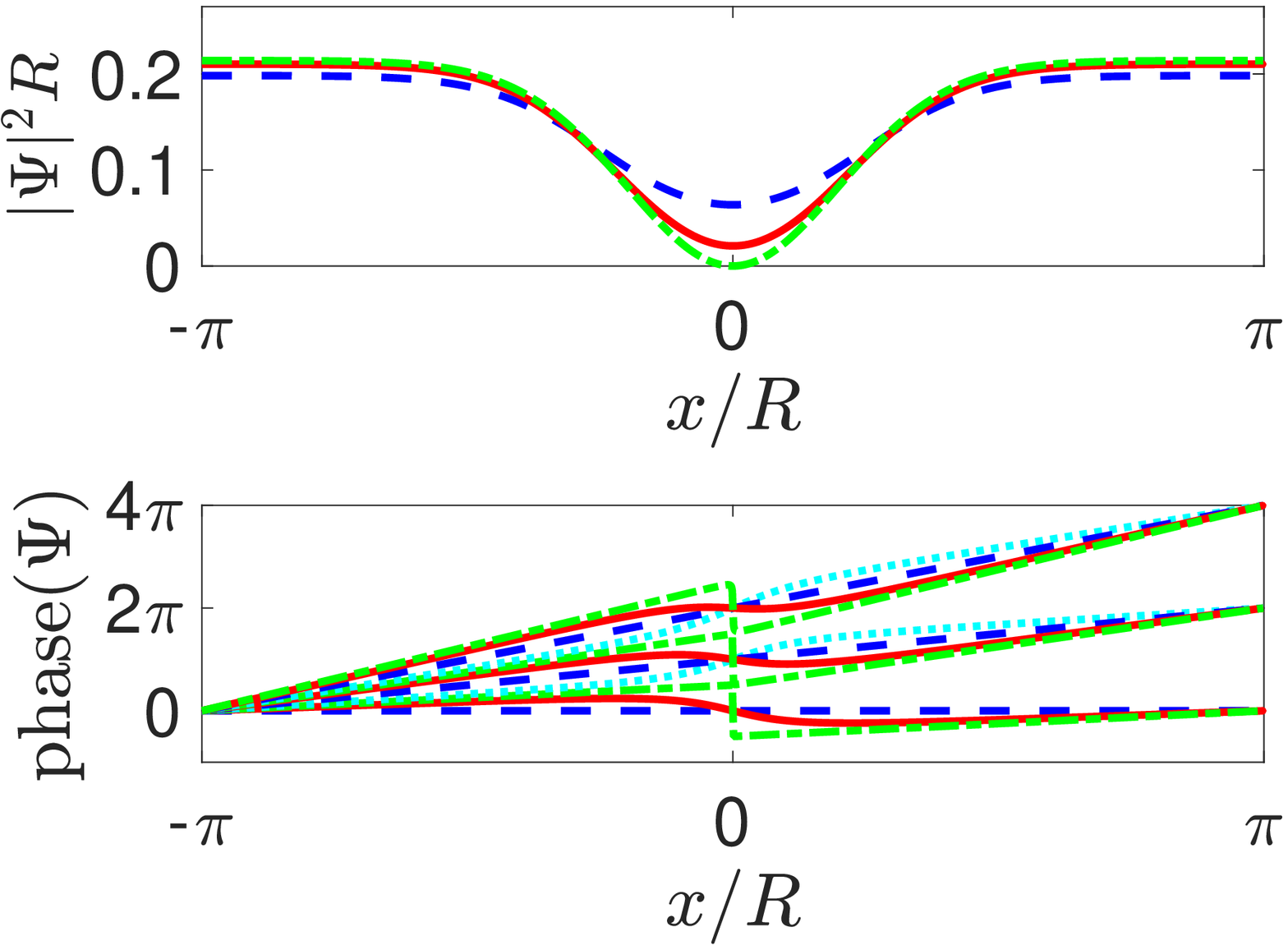}
\includegraphics[width=4cm,height=4.cm]{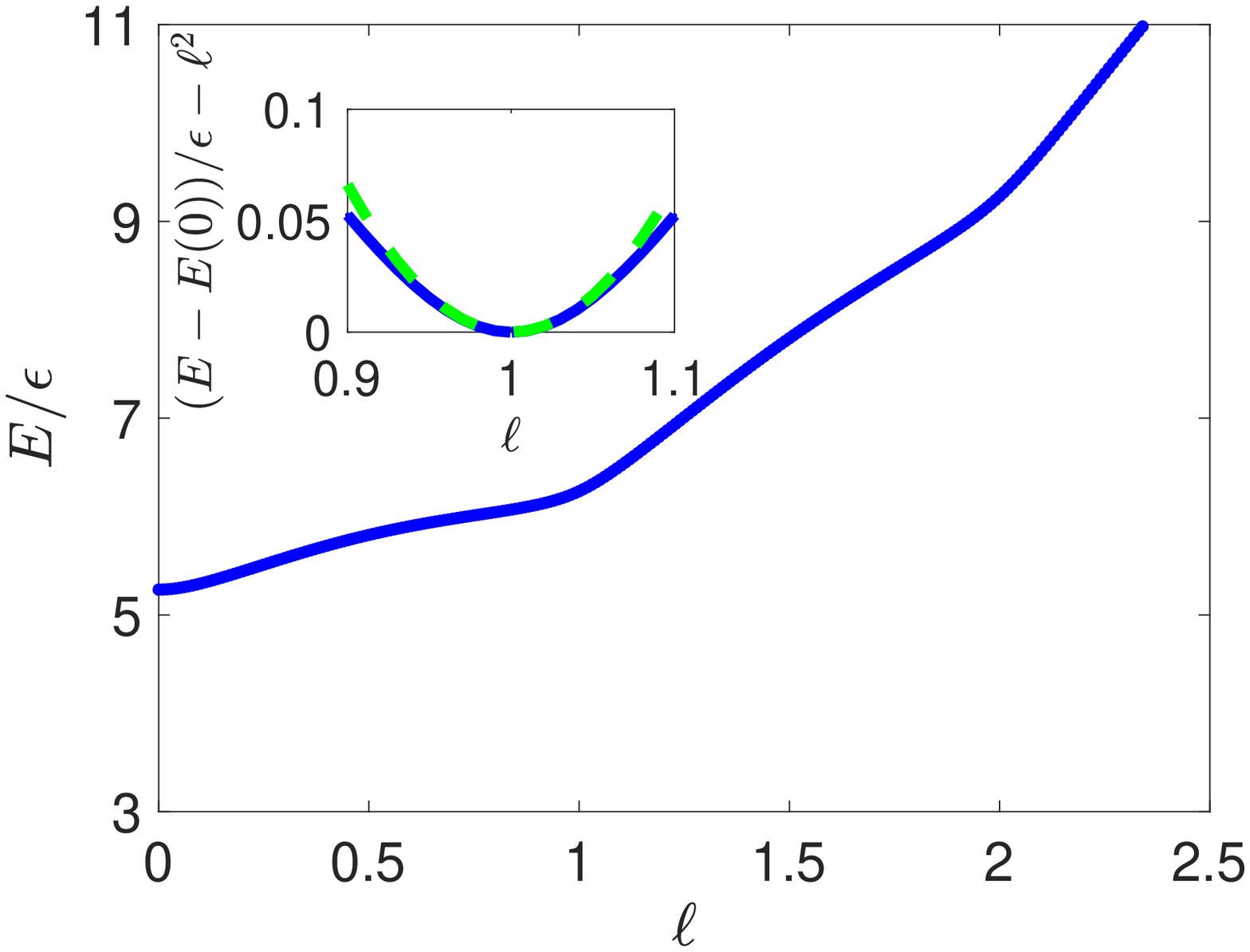}
\includegraphics[width=4cm,height=4.cm]{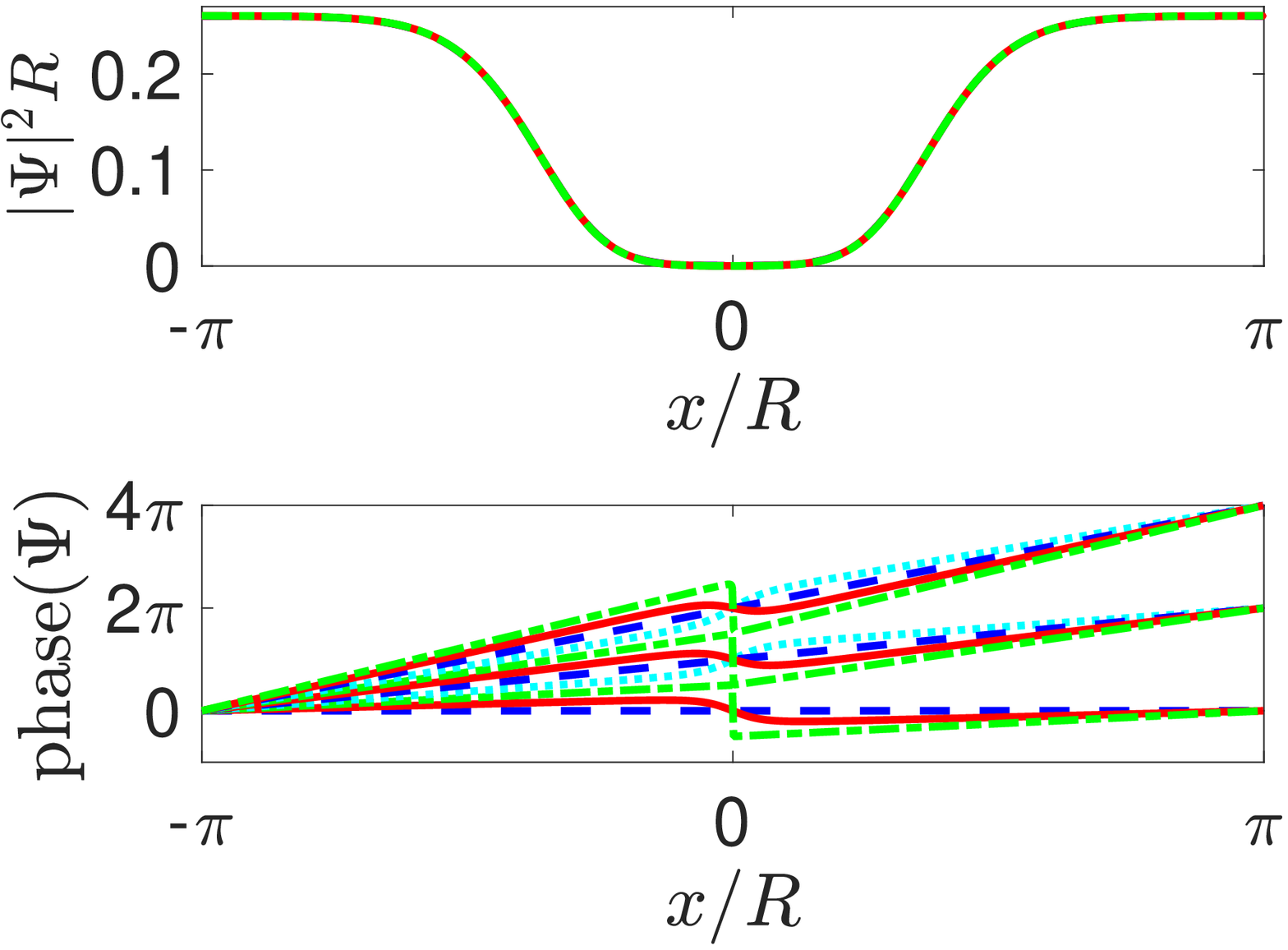}
\includegraphics[width=4cm,height=4.cm]{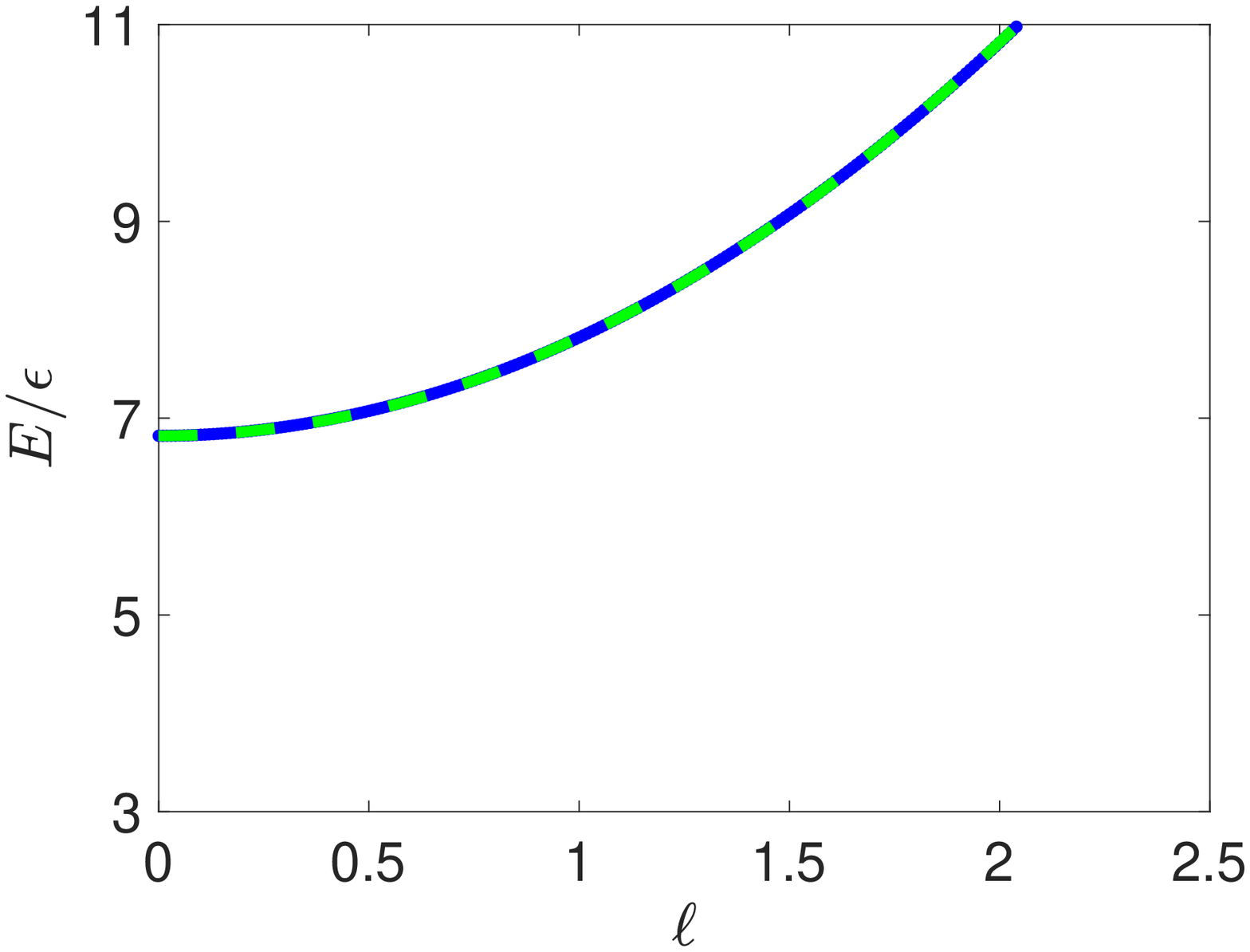}
\caption{(Color online) The density, the phase (left plots), and the dispersion relation (right plots), 
for $\delta = 7.5$, $w_0/R = 0.6$, and $V_0/\epsilon=0$ (top), $V_0/\epsilon=4$ (second from the top), 
$V_0/\epsilon=8$ (third from the top), and $V_0/\epsilon=32$ (bottom). In the plots of the density the 
cases $\ell = 0, 1$, and 2 are denoted as blue dashed lines, $\ell = 0.25, 0.75, 1.25, 1.75$, and 2.25 
as red solid, and $\ell = 0.5, 1.5$, and 2.5 as green dashed-dotted. In the plots of the phase, the cases 
$\ell = 0, 1$, and 2 are denoted as blue dashed lines, $\ell = 0.25, 1.25$, and 2.25 as red solid, $\ell 
= 0.75$ and 1.75 as cyan dotted, and $\ell = 0.5, 1.5$, and 2.5 as green dashed-dotted. The green dashed 
lines in the top right plot come from Eq.\,(\ref{disprell}), for $q = 0$, 1, and 2. The green dashed 
parabolas in the insets on the left of the second and the third right plots come from Eq.\,(\ref{enlimm2}), 
for $q = 1$. The inset plot at the bottom right of the second right plot focuses around $\ell = 0.9$, where 
there is a local minimum of $E(\ell)$ at $\ell \approx 0.9155$ (vertical black line). The green dashed 
parabola in the bottom right plot is $E(0)/\epsilon + \ell^2$ and describes rigid-body rotation.}    
\label{fig1}
\end{figure}

\begin{figure}
\includegraphics[width=4cm,height=5cm,angle=-0]{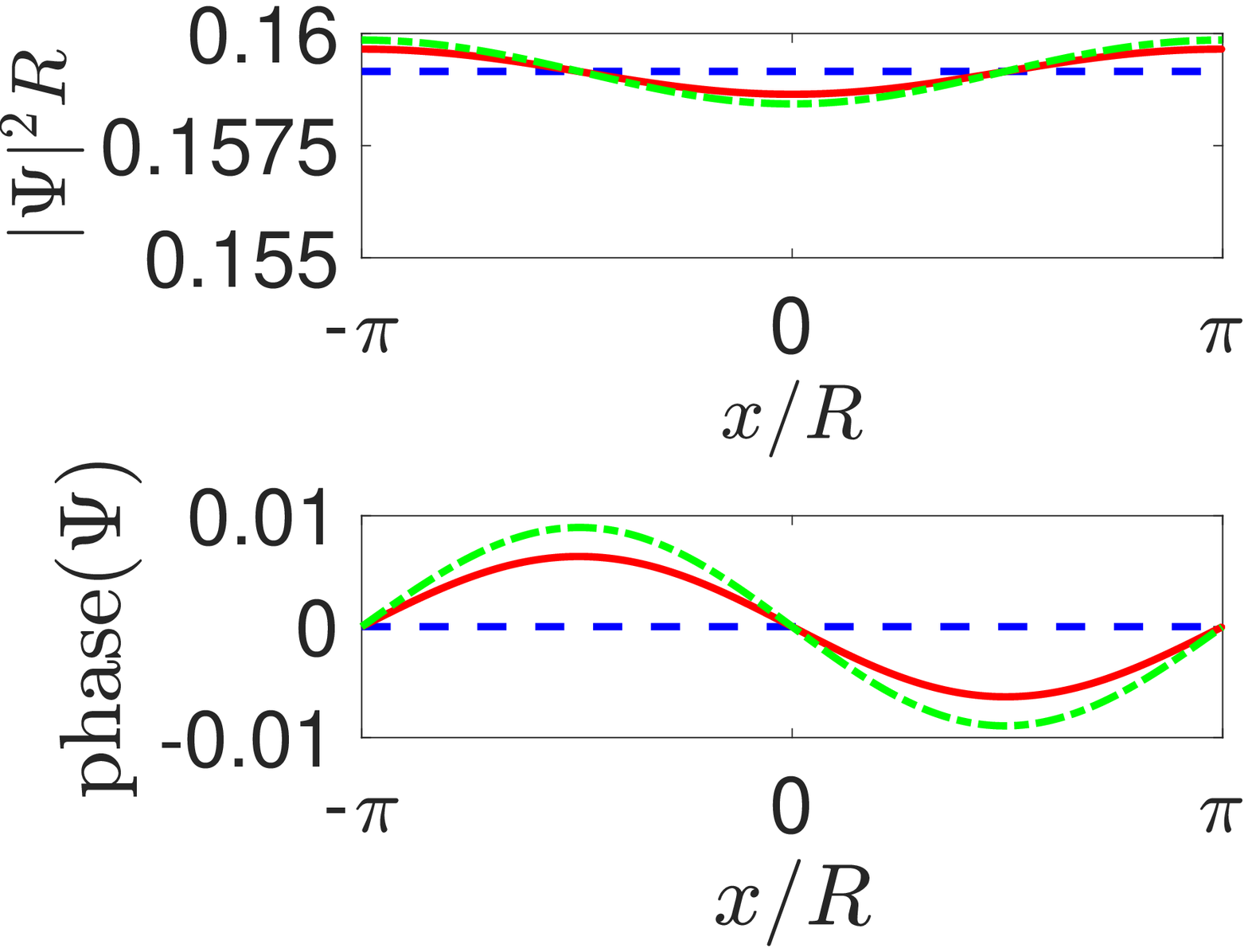}
\includegraphics[width=4cm,height=5cm,angle=-0]{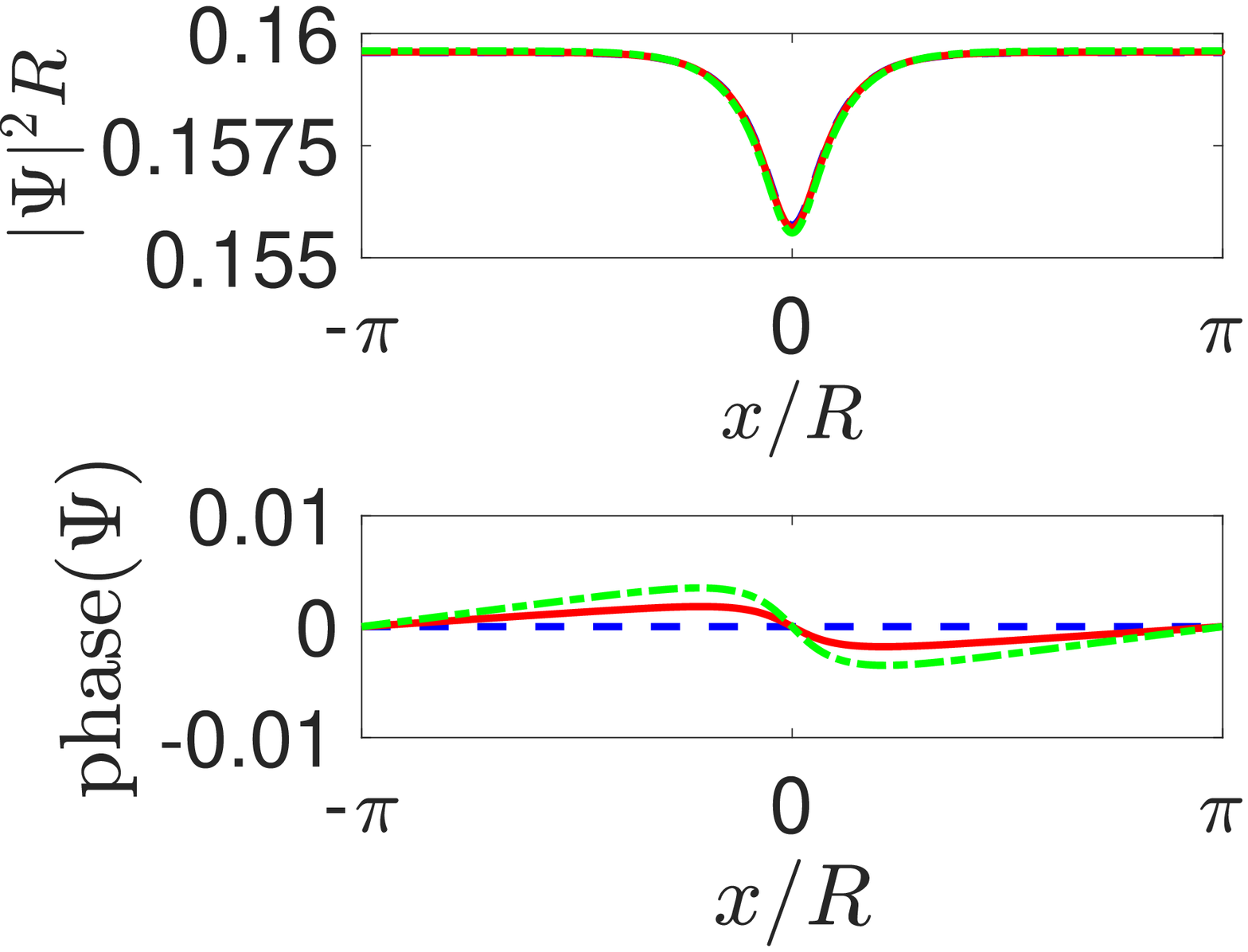}
\caption{(Color online) The density (upper) and the phase (lower) of the order parameter, for $\ell = 0$ (blue, 
dashed curve), $\ell = 1 \times 10^{-5}$, (red, solid curve), and $\ell = 2 \times 10^{-5}$ (green, dashed-dotted
curve), with $\delta = 7.5$, $w_0/R = 0.1$, and $V_0/\epsilon = 0$ (left), $V_0/\epsilon = 0.5$ (right).}
\label{fig2}
\end{figure}  

\section{Dispersion relation in the presence of a potential}

We turn now to the effect of an external potential. First of all, we stress that the other
relevant energy scale is the chemical potential, which, in turn, is set by $n_0 g$. Even if the potential 
is weak, for sufficiently small values of $|\ell-q|$, increasing $\ell$ in this regime, changes only the 
phase of the order parameter, with the density remaining unchanged, and thus the only change takes place 
in the superfluid velocity $v$. This effect is seen clearly in Fig.\,2. 

Returning to our problem, if we set $\Psi(x) = |\Psi(x)| e^{i \phi(x)}$ in Eq.\,(\ref{gpesss}), then since 
the superfluid velocity $v(x)$ is equal to $\hbar \phi'(x)/M$, we get after integrating the 
continuity equation, 
\begin{eqnarray}
  \frac {\partial n} {\partial t} + \frac {\partial} {\partial x} (n v) = 0,  
\end{eqnarray}
where $n = N |\Psi(x)|^2$ is the density, that
\begin{eqnarray}
   v(x) = u + \frac C {n(x)} = \Omega R + \frac C {n(x)},
   \label{cccc}
\end{eqnarray}
where $C$ is the constant of integration. Taking the integral 
\begin{eqnarray}
  \int_{-\pi R}^{\pi R} v(x) dx = \frac {\hbar} M [\phi(\pi R) - \phi(- \pi R)].
  \label{dddd}
\end{eqnarray}
Assuming that the phase difference in the above equation is equal to $2 \pi q$, it follows from Eqs.\,(\ref{cccc})
and (\ref{dddd}) that 
\begin{eqnarray}
 v(x) = \Omega R \left( 1 - 2 \pi R \frac {\kappa} {n(x)} \right) + \frac {2 \pi \kappa q \hbar} {M n(x)},
\end{eqnarray}
where $\kappa^{-1} = N \int n^{-1}(x) dx$. Turning to the angular momentum,
\begin{eqnarray}
    \ell \hbar = M R \int n v \, dx = M \Omega R^2 \left( 1 - \frac {\kappa} {\kappa_0} \right) 
    + \frac {\kappa} {\kappa_0} q \hbar, 
\nonumber \\
\label{allim2}
\end{eqnarray} 
with $\kappa_0 = (2 \pi R)^{-2}$ being the value of $\kappa$ for the homogeneous state. In addition, the 
kinetic energy is, 
\begin{eqnarray}
\frac {K_q(\ell)} {\epsilon} &=& \frac {(\ell - \ell_q)^2} {1 - \kappa/\kappa_0} 
+ q^2 \frac {\kappa} {\kappa_0}
 = \ell^2 + \frac {(\ell - q)^2} {\kappa_0/\kappa - 1},
\label{enlimm2}
\end{eqnarray}
where $\ell_q = (\kappa/\kappa_0) q$. Therefore, in the regime mentioned above (of $\ell \approx q$), since 
the density is constant in this range of $\ell \approx q$, the interaction energy is also constant. As a 
result the excitation energy of the system is purely kinetic and it scales quadratically with $\ell$, with 
an effective moment of inertia $I = M R^2 \left( 1 - \kappa/\kappa_0 \right)$ \cite{miner}. 

The last equality of Eq.\,(\ref{enlimm2}) shows the agreement with Bloch's theorem, with the periodic function 
$e(\ell)$ which appears in Eq.\,(\ref{drel}) being $(\ell - q)^2(\kappa_0/\kappa - 1)$ (for $\ell \approx q$). 
We stress that the sum of the interaction and of the potential energies of the system, which we denote as $E(0)$, 
is constant for $\ell \approx q$. Another important result that follows from Eq.\,(\ref{enlimm2}) is that the 
center of the parabolas is not at integer values of $\ell$, but rather at the values of $\ell = \ell_q$. 

The interesting quantity which appears in the analysis presented above is $\kappa$, which is associated with 
the (angular-momentum-independent) density distribution of the gas, for sufficiently small values of $|\ell - q|$. 
This may be evaluated numerically. We can also get an analytic estimate for $\kappa$ via an approximate expression 
for $n(x)$ in the Thomas-Fermi limit. For example, for an external potential of the form of Eq.\,(\ref{expot}), 
in the limit where its width $w_0$ is larger than the coherence length $\xi$, and also when $V_0$ is larger 
than $\hbar^2/(M w_0^2)$, then the kinetic energy in Eq.\,(\ref{gpe}) may be neglected, and 
\begin{eqnarray}
  n(x) \approx n_0 \left( 1 +  \frac 1 {\sqrt{2 \pi}} \frac {w_0} R \frac {V_0} {n_0 g} 
  - \frac {V_0} {n_0 g} e^{-x^2/(2 w_0^2)} \right). 
  \label{densw}
\end{eqnarray}
From Eq.\,(\ref{densw}) it follows that
\begin{eqnarray}
 \ell_q \approx q \left[ 1 - \frac 1 {2 \sqrt{\pi}} \frac {w_0} {R} \left( \frac {V_0} {n_0 g} \right)^2 \right].
  \label{ellq}
\end{eqnarray}

As $\ell$ deviates from $\ell = \ell_q$, the quadratic dependence of the energy on $\ell$ makes this kind of excitation 
energetically unfavourable. As a result, unless the potential is very strong, as $\ell$ increases there is a change 
in the nature of the excitation, with the system behaving as in the case where the potential is absent, i.e., as in 
the axially-symmetric case. Thus, one may identify a critical value of $\ell$, which we denote as $\ell_{c,q}$. For 
values of $\ell \lesssim \ell_{c,q}$ the dispersion relation is determined by the presence of the external 
potential, and is quadratic in $\ell$ in this regime, with Eq.\,(\ref{enlimm2}) being valid. On the other hand, 
for $\ell \gtrsim \ell_{c,q}$ the dispersion relation is analogous (roughly) to the one where the potential is 
absent [and, for sufficiently small values of $|\ell-q|$, it is linear in $\ell$, according to Eq.\,(\ref{disprell})]. 

We estimate the value of $\ell_{c,q}$ by equating the slope of the dispersion relation of Eq.\,(\ref{disprell}) 
(in the case where there is no potential) with the slope of the quadratic dispersion relation, Eq.\,(\ref{enlimm2}), 
thus finding
\begin{eqnarray}
  \ell_{c,q} \approx q + \frac 1 2 \sqrt{2 \delta + 1} \left(1 - \frac {\kappa} {\kappa_0} \right).
  \label{ellc}
\end{eqnarray}
Using the approximate expression of Eq.\,(\ref{densw}),
\begin{eqnarray}
  \ell_{c,q} \approx q + \sqrt{2 \delta + 1} 
  \frac 1 {4 \sqrt{{\pi}}} \frac {w_0} {R} \left( \frac {V_0} {n_0 g} \right)^2.
\end{eqnarray}
An important observation in the above equation is that the terms which are of order $({w_0}/{R}) ({V_0}/{n_0 g})$ 
cancel. As a result, the behaviour is the same for both signs of $V_0$, i.e., for attractive/repulsive potentials. 

Turning back to our numerical results, in Fig.\,1, we consider a potential of the form of Eq.\,(\ref{expot}), 
with $V_0 > 0$ (i.e., repulsive), with an increasing strength $V_0$. As a result, even for $\ell = 0$, we see 
that the density of the gas is inhomogeneous. Other than that, the behaviour of the density is roughly the same 
as in the case with $V_0 = 0$. As $\ell$ increases from zero, the minimum of the density decreases, all the way 
up to $\ell = 1/2$. Then, the minimum of the density increases up to $\ell = 1$.

In addition, the dispersion relation $E(\ell)$ has a quadratic dependence on $\ell-q$, as long as $V_0 \neq 0$,  
for sufficiently small values of $|\ell-q|$, in agreement with Eq.\,(\ref{enlimm2}). Furthermore, the center of 
these parabolas is located at the values $\ell_q$, Eq.\,(\ref{ellq}). In addition, $\Omega$ no longer has 
discontinuities, but rather it vanishes at the values of $\ell = \ell_q$ (see Fig.\,4). 

In the specific example of the second set of data in Fig.\,1 we see that the center of the local minimum of 
the dispersion relation $E(\ell)$ is at $\ell \approx 0.9155$, which corresponds to $\ell_{q=1}$ in 
Eq.\,(\ref{enlimm2}). For the parameters of this data $\kappa \approx 0.02460$, which implies that $\ell_{q=1} 
\approx 0.9712$, with a difference of roughly 6\% from the numerical result. This local minimum in $E(\ell)$ 
is also seen as a node in its derivative, i.e., in $\Omega$, shown in Fig.\,4. Finally, the (trivial) time 
evolution of this static and gray solitary wave is seen in the upper plot of Fig.\,5. In the same figure we 
also show in the lower plot the case of an attractive potential, for the same value of $\ell \approx 0.9155$. 

As $V_0$ increases further (third and bottom plots in Fig.\,1), the minimum of the density at $\ell = 0$ 
decreases. In the extreme case of the bottom plot of Fig.\,1, with the largest value of $V_0$, the potential 
is sufficiently strong, so that, even in the absence of any rotation, the density vanishes within some range 
along the ring. In this case the motion of the system resembles solid-body rotation, as the density distribution 
of the atoms is independent of the angular momentum. Furthermore, the phase is linear in the regions where the 
density is constant, i.e., far away from the density depression, varying only in the region where the density 
is vanishingly small. Finally, the dispersion relation is quadratic in $\ell$, for all values of $\ell$.

\begin{figure}[t]
\includegraphics[width=4cm,height=4cm,angle=-0]{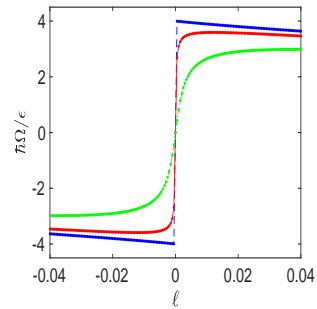}
\caption{(Color online) The function $\Omega = \Omega(\ell)$, for $\delta = 7.5$, $w_0/R = 0.1$, and $V_0/\epsilon 
= 0$, (blue, outer curve), 1 (red, middle curve), and 4 (green, inner curve).}
\label{fig3}
\end{figure} 

\begin{figure}[h]
\includegraphics[width=4cm,height=4.cm]{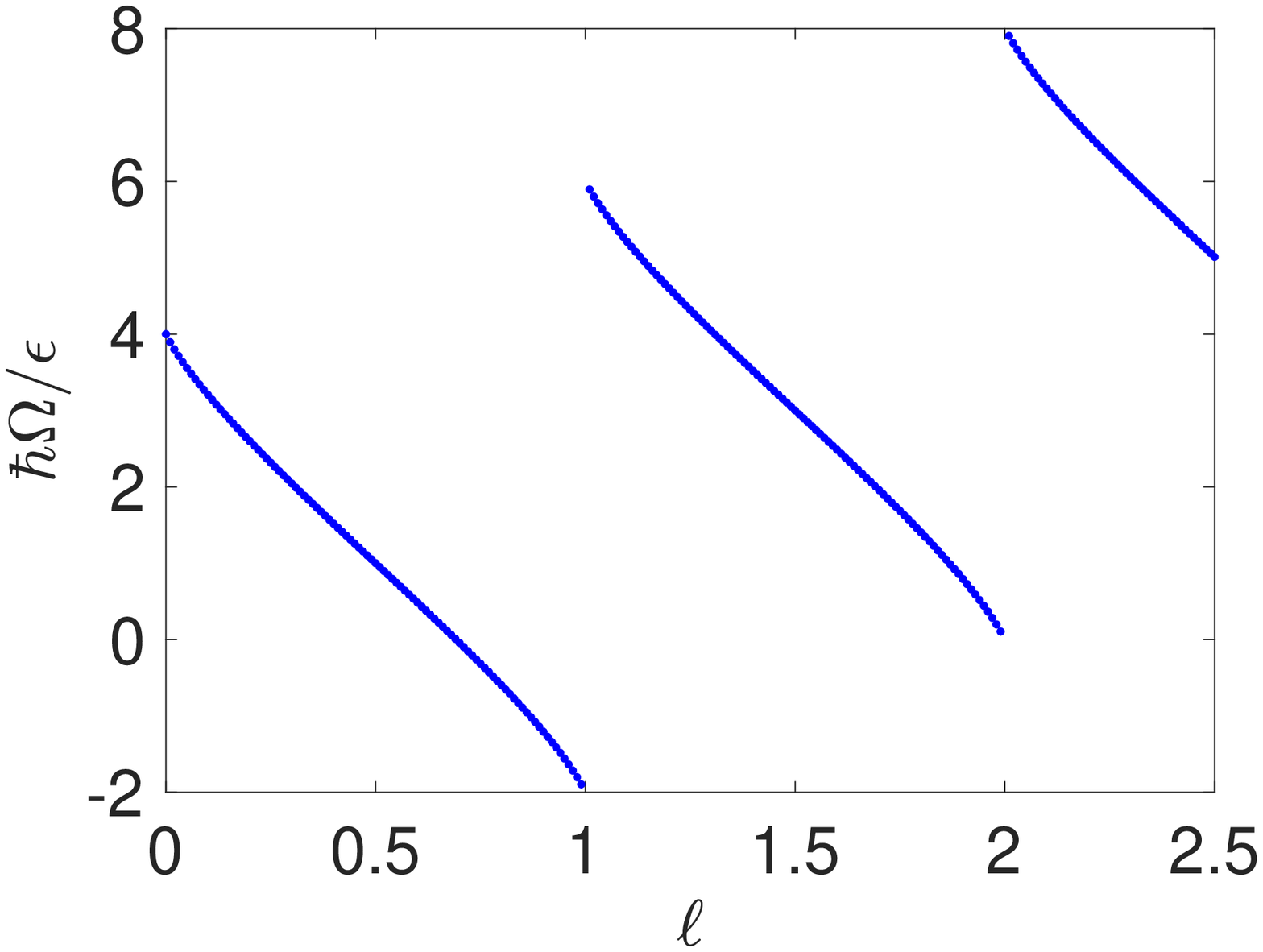}
\includegraphics[width=4cm,height=4.cm]{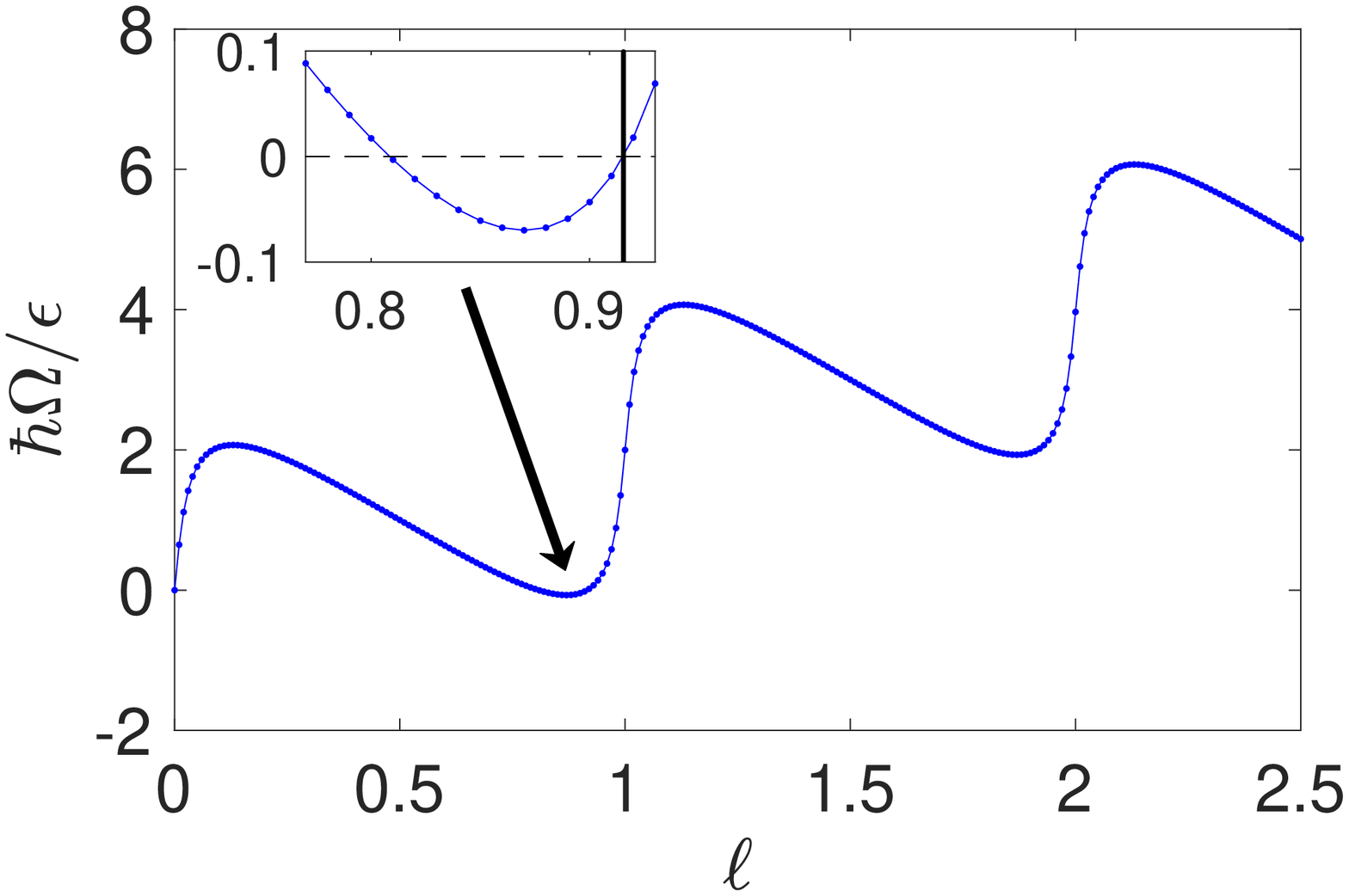}
\includegraphics[width=4cm,height=4.cm]{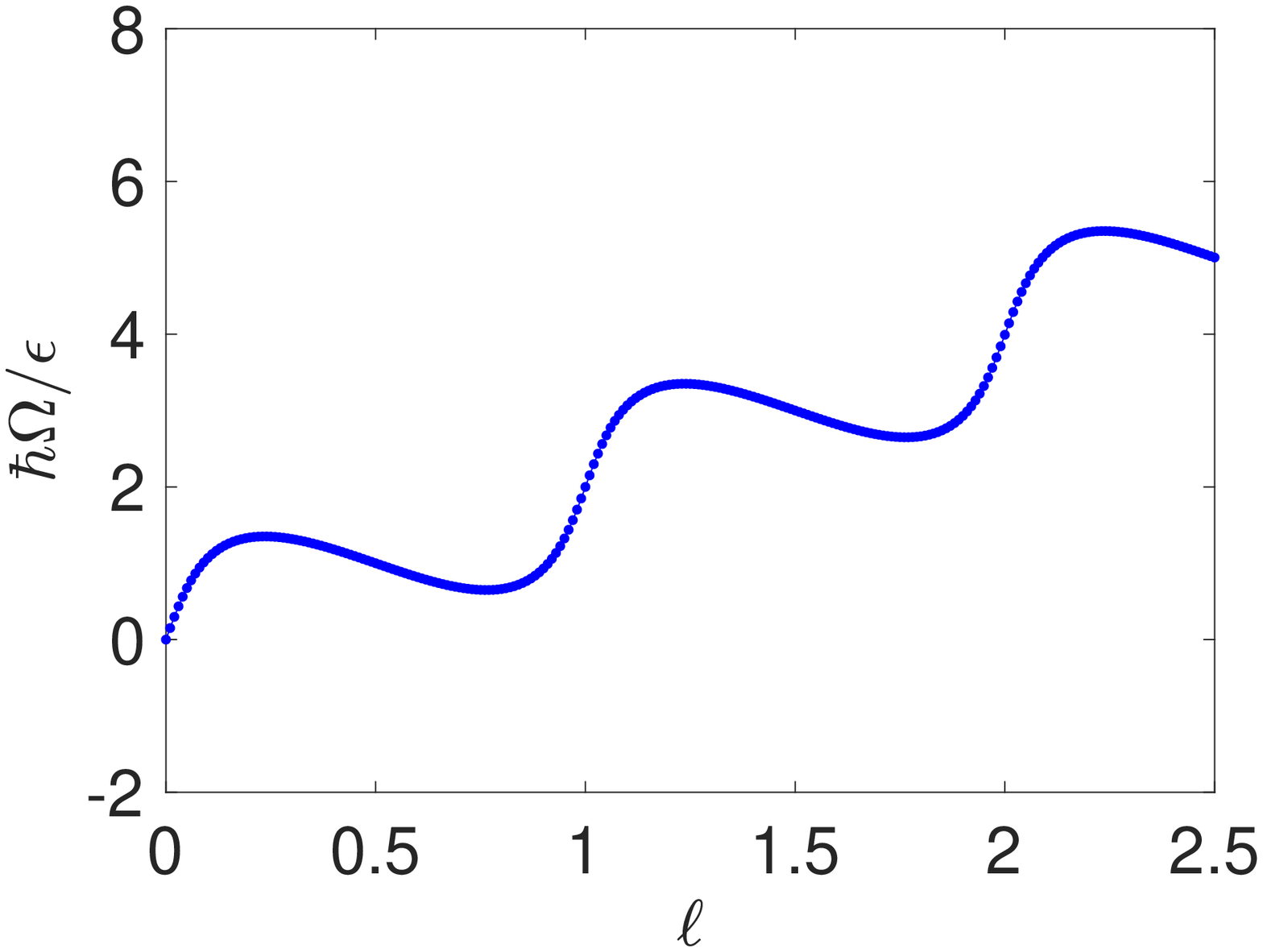}
\includegraphics[width=4cm,height=4.cm]{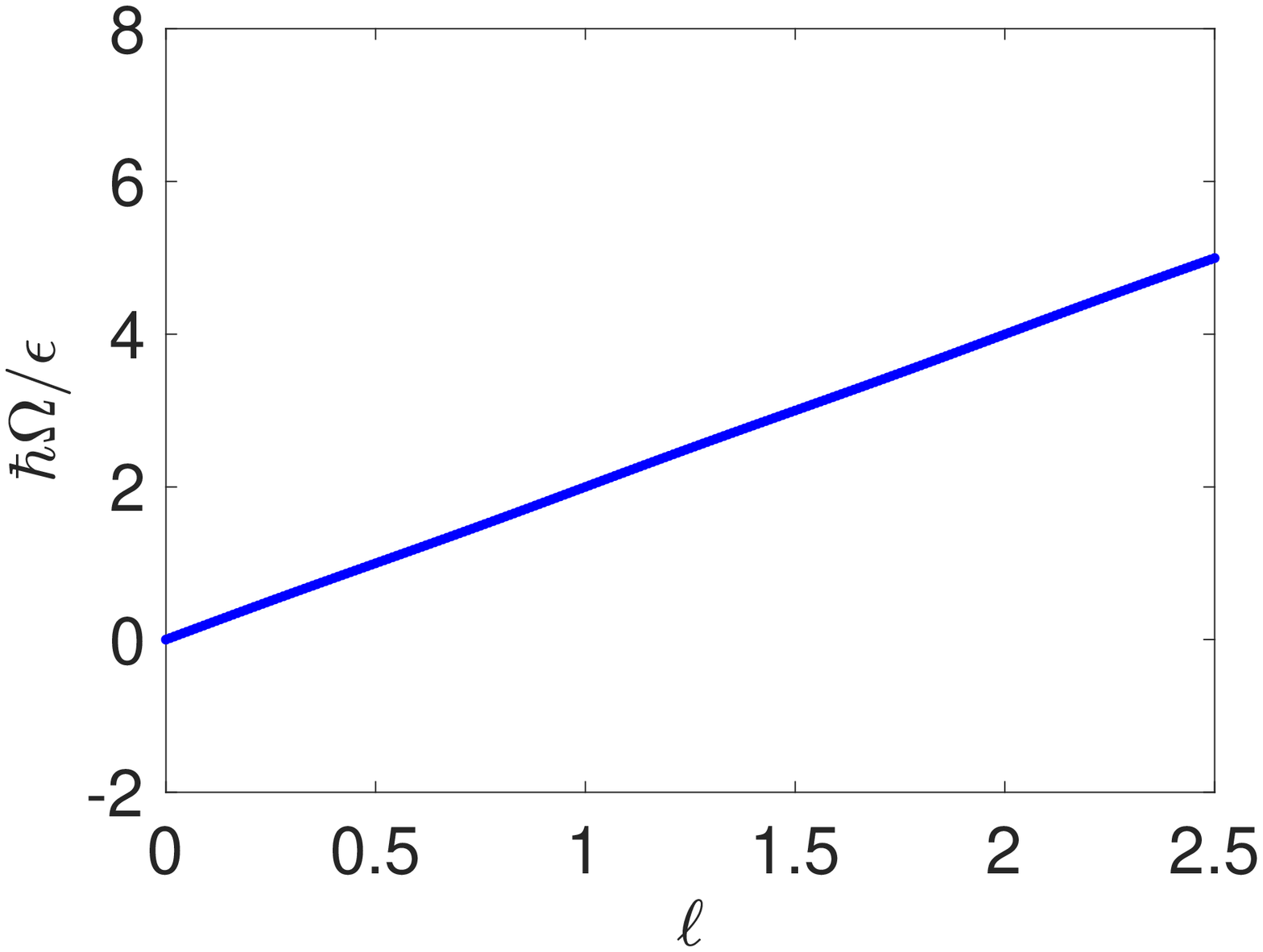}
\caption{(Color online) The function $\Omega = \Omega(\ell)$, for $\delta = 7.5$, $w_0/R = 0.6$, and
$V_0/\epsilon=0$ (upper left), $V_0/\epsilon= 4$ (upper right), $V_0/\epsilon=8$ (lower left), and 
$V_0/\epsilon = 32$ (lower right). The inset in the upper right plot shows the two points where 
$\Omega$ vanishes, where the node on the right is at $\ell \approx 0.9155$ (vertical black line). 
These two points may also be seen in the second right plot of Fig.\,1 (as extrema in the dispersion 
relation).}    
\label{fig4}
\end{figure}

\begin{figure}[h]
\includegraphics[width=6cm,height=6cm,angle=-0]{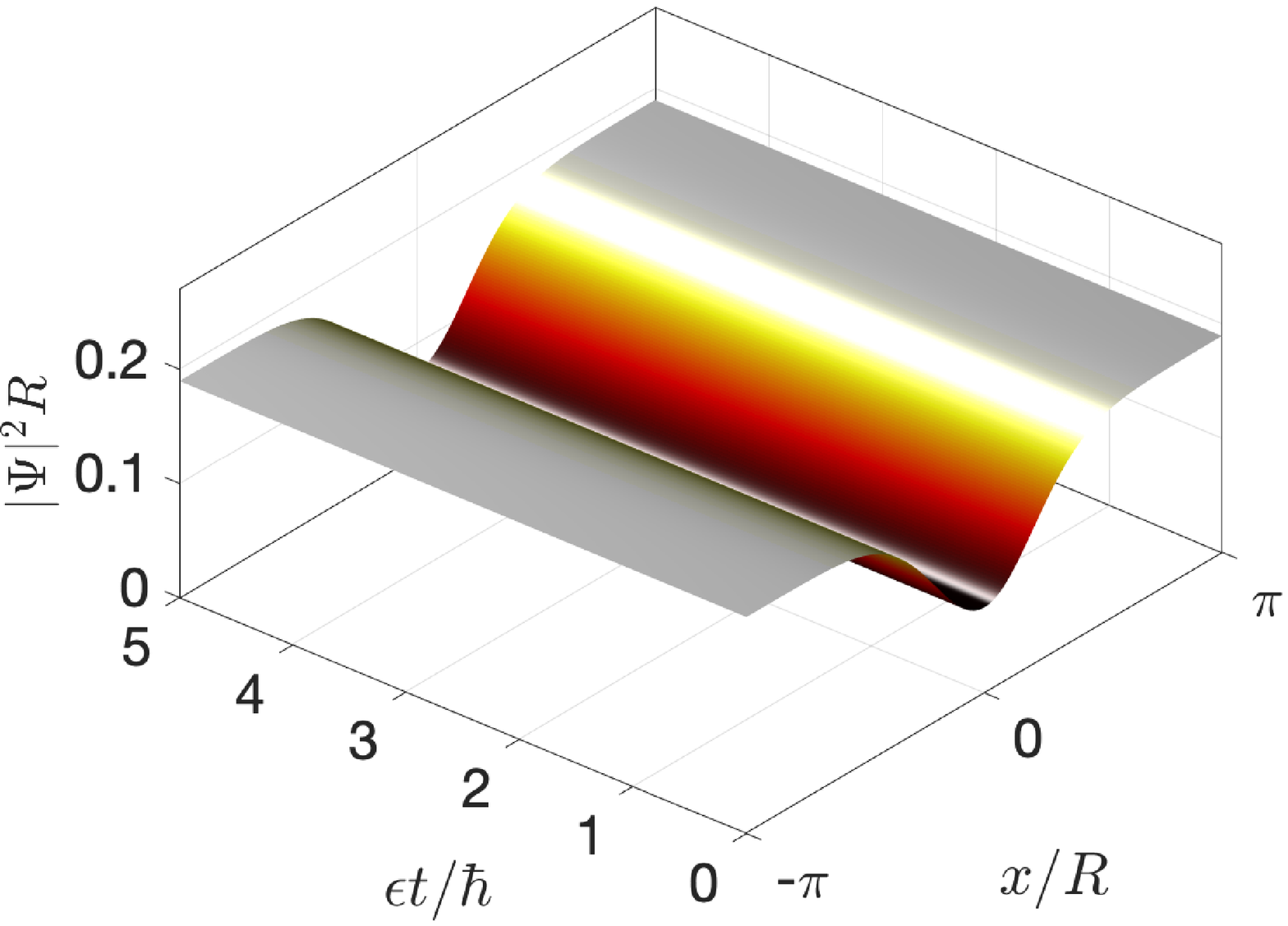}
\includegraphics[width=6cm,height=6cm,angle=-0]{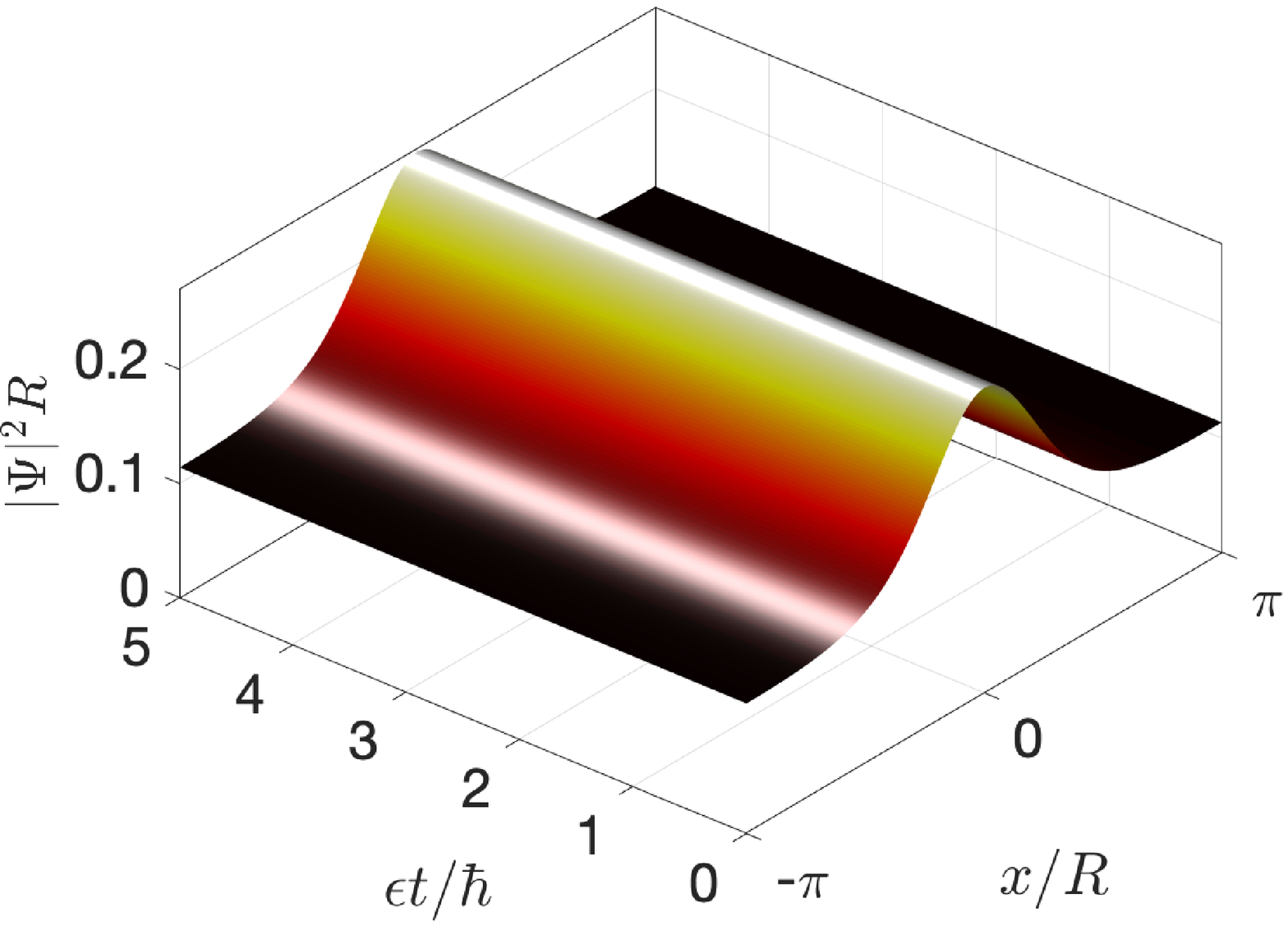}
\caption{(Color online) Upper: The time evolution of the static solution (i.e., with $\Omega \simeq 0$), for 
a repulsive potential, $V_0/\epsilon = 4$, for the value of $\ell \approx 0.9155$ (shown also in Figs.\,1 
and 4), $\delta = 7.5$, and $w_0/R = 0.6$. Lower: Same as above, for an attractive potential, with 
$V_0/\epsilon = - 6.532$.}
\label{fig5}
\end{figure} 

\section{Effect of the potential on the observables}

One of the main results of the present study is the angular velocity $\Omega$. In the axially-symmetric problem 
($V_0 = 0$), the dispersion relation has discontinuities in its slope at the integer values of $\ell$. As a result, 
the function $\Omega=\Omega(\ell)$ has discontinuous jumps, while $\Omega$ varies in the range between $\Omega = 
(2q - \sqrt{2 \delta + 1}) \epsilon/\hbar$, and $\Omega = (2q + \sqrt{2 \delta + 1}) \epsilon/\hbar$, as discussed 
in Sec.\,IV. 

Figure 3 demonstrates the effect of the external potential on $\Omega$. When there is no external potential, we 
see the expected discontinuities, however these disappear in the additional two curves, which correspond to two 
nonzero values of $V_0$. Furthermore, the width in $\ell$ over which we have the quadratic dependence is on the 
order of $\ell_{c,q}$, Eq.\,(\ref{ellc}), which increases with increasing $V_0$. 

Figure 4 shows $\Omega$ in a wider range of values of $\ell$, for some representative values of $V_0$, including 
the case $V_0 = 0$, in the first plot. Here we see clearly the quasi-periodic behaviour due to Bloch's theorem 
and also the discontinuities described earlier. Another general characteristic of $\Omega$ is that it is a 
decreasing function of $\ell$ because the curvature of the dispersion relation is negative (for effectively 
repulsive interatomic interactions). 

In the presence of an external potential, the above picture changes. First of all, the discontinuities in $\Omega 
= \Omega(\ell)$ disappear, while $\Omega$ always vanishes for $\ell = 0$. Furthermore, $\Omega$ still has a 
quasi-periodic behaviour in $\ell$, due to Bloch's theorem, with a period which is still equal to unity. Unless 
the potential is very strong (compared with the interaction energy), $\Omega$ increases from zero with increasing 
$\ell$, up to roughly $\ell_{c, q=0}$. As $\ell$ increases further, $\Omega$ starts to drop, up to some value of 
$\ell$ which is of order $1 - \ell_{c,q=1}$, increasing again up to some value, which is on the order of $1 + 
\ell_{c, q=1}$, with this continuing quasi-periodically, as seen in the second and the third plots of Fig.\,4.

In addition, as the strength of the potential increases, $\ell_{c, q}$ increases, too. Eventually, when the potential 
becomes sufficiently strong, the dispersion relation becomes parabolic for all values of $\ell$ and $\Omega$ becomes
trivially a straight line (last plot in Fig.\,4). Therefore, as $V_0$ increases, we have a gradual transition of the 
motion of the system to classical, solid-body-like rotation.

Of particular importance in this plot are the points where $\Omega$ vanishes, with $\ell \neq 0$. When the 
potential is not very strong and the interaction is sufficiently strong, this happens for two values of $\ell$
in each interval of quasi-periodicity (which extends roughly between $\ell =  q$ and $\ell = q+1$). The one is 
already present, even in the absence of an external potential and corresponds to a static solitary wave which 
is also ``grey" \cite{aaa}. In the presence of a potential, when this is not very strong, this point is shifted 
slightly. More importantly, a second node in $\Omega$ may appear (depending on the value of the parameters), 
roughly at the values of $\ell = q - \ell_{c,q}$. While the first node corresponds to a maximum of the dispersion 
relation, the second corresponds to a minimum. Such an example is shown in Fig.\,1 (see also Figs.\,4 and 5).

While in all the solutions we have found there is no relative motion between the solitary wave and
the potential, at these points both the potential, as well as the wave are static, in the lab frame. On the other 
hand, both the circulation, as well as the superfluid velocity are nonzero. Figure 5 shows the real-time evolution 
of such two states. As seen in this plot, the density is inhomogeneous and does not change in time, as expected. 
Remarkably, under these conditions, the fluid passes through the ``obstacle" potential without any change, with 
the gas obviously having an inhomogeneous density distribution. Thus, in a sense this is a realization of a 
reflectionless potential. This is also a situation where we have a persistent current, although the density of 
the gas is inhomogeneous (as opposed to the case of axial symmetry, where persistent currents show up for a 
homogeneous density distribution). 

Another immediate consequence of Fig.\,4 is that, for a fixed value of $\Omega$ there may be many states with different 
$\ell$. Depending on how one performs an experiment, he may end up in any of these states.

\section{Experimental relevance}

In making contact with experiments, let us consider, for example, the parameters of Ref.\,\cite{hysteresis}. 
In this experiment an annular potential was used, with trap frequencies $\omega_1 \approx 472$ Hz and $\omega_2
\approx 188$ Hz. In addition, the chemical potential $\mu/\hbar$ was $\approx 2 \pi \times 1.7$ kHz. Under 
these conditions the assumption of (quasi) one dimensional motion -- which is made in the present study -- 
is not satisfied. In order to achieve this condition, one would have to make the oscillator quantum of energy 
associated with the transverse degrees of freedom larger than the chemical potential, and thus freeze the 
motion of the atoms in this direction.

In the experiment of Ref.\,\cite{hysteresis} $N \approx 4 \times 10^5$ and the ring radius $R \approx 19.5$ 
$\mu$m, thus $\epsilon/\hbar \approx 3.6$ Hz, while $\delta \approx 1500$, for a value of the scattering length 
$\approx 28$ \AA \, for $^{23}$Na atoms. With these numbers, the energy associated with the excitation of the 
center of mass is negligible, while the typical scale of $\Omega$ is on the order of $\mu/\hbar \approx 10$ kHz. 
Reducing the atom number, or increasing the trap frequencies $\omega_1$ and $\omega_2$ would eventually drive the 
system in the quasi-one-dimensional limit. In order for the energy associated with the excitation of the center 
of mass motion to be non-negligible, $\delta$ would have to be $\gtrsim 1$. In this case the typical value of 
$\Omega$ would be on the order of 1-10 Hz. 

Finally, about the length scales, in Ref.\,\cite{hysteresis} the coherence length $\xi$ was roughly $R/\sqrt{\delta}$, 
i.e., $\approx 0.5$ $\mu$m, for $\delta = 1500$. This should be compared with the width of the potential $w_0$, 
which was roughly 6 $\mu$m in the experiment of Ref.\,\cite{hysteresis} and therefore in this case $\xi < w_0 < R$. 
Finally, for $\delta \gtrsim 1$, then $\xi \lesssim R$, i.e., in this limit we have that $w_0 \lesssim \xi \lesssim R$.

\section{Summary and conclusions}

In the present problem we considered a Bose-Einstein condensed gas of atoms, confined in a narrow torus/annulus,
which we approximated with a ring potential, with an effective repulsive interaction between the atoms. In this 
system, when there is no external potential, the excitation spectrum has certain characteristics, which determine its 
rotational response. For sufficiently small values of the angular momentum, the dispersion relation scales linearly 
with the angular momentum, while for larger values its curvature is negative. Furthermore, there are discontinuities 
in the slope of the dispersion relation -- and as a result in $\Omega = \Omega(\ell)$ -- at all integer values of the 
angular momentum $\ell \hbar$. 

In the presence of an external potential, which is assumed to rotate with the same angular velocity as the wave, 
there are interesting observations, as described in Secs.\,V and VII. First of all, for small values 
of the angular momentum the dispersion relation becomes quadratic, with a positive curvature. In this regime the 
rotation resembles that of a solid-body. As the angular momentum increases, eventually it becomes energetically 
favourable for the system to carry its angular momentum via vortex excitation, as in the case where the external 
potential is absent. Thus, there is a transition from the one kind of motion to the other. 

One important consequence of the symmetry-breaking potential is that the function $\Omega = \Omega(\ell)$ is 
continuous, as seen in Fig.\,4. In this figure we also observe that when the strength of the external potential 
becomes sufficiently strong, the system behaves as a solid-body and $\ell = \ell(\Omega)$ becomes trivial, with 
$\Omega$ increasing linearly with $\ell$. 

We have thus identified a transition as the strength of the external potential increases. In this transition we 
observe a ``hybrid" behaviour between solid-body-like excitation and vortex excitation. Depending on the value of 
the relevant parameters, the one dominates over the other. 

In addition the system we have studied may support persistent currents, with an inhomogeneous density distribution 
(which is essentially a realization of a ``reflectionless potential), as seen in Fig.\,5. 

The results presented above are not only interesting from a theoretical point of view, but they may also be useful 
in the field of atomtronics, or, more generally in the field of atom manipulation. The large degree of tunability   
that we have on these systems (and, in particular, on the strength of the stirring potential, which plays a crucial
role) may allow experimentalists to design devices, which will make use of the superfluid properties that we 
investigated. Finally, it is an open question whether our results persist in a wider annulus/torus, when the
assumption of quasi-one-dimensional motion of the atoms is no longer valid.

\acknowledgements M. \"{O}. acknowledges partial support from ORU-RR-2020/2021 and from the Magnus Bergvalls 
foundation. G.M.K. wishes to thank M. Magiropoulos and J. Smyrnakis for useful discussions.


\begin{thebibliography}{99}
\bibitem{remark} Another way to argue about the conservation of the angular momentum is the 
following. The time derivative of the expectation value of the operator of the angular momentum 
$\hat{\ell} = - i \hbar R d/dx$, $\langle \hat{\ell} \rangle$, is proportional to $\langle dV/dx 
\rangle$, which follows from Ehrenfest's theorem. When $V(x)$ is an even function of $x$, then 
$dV/dx$ is an odd function, while the density $n(x) = N |\Psi(x)|^2$ is even, and thus for the 
solutions of Eq.\,(\ref{gpesss}) $\langle dV/dx \rangle$ vanishes trivially. As a result, 
$\langle \hat{\ell} \rangle$ is constant. Actually, even if $V(x)$ does not have any symmetry, 
the expectation value $\langle dV/dx \rangle$ in any converged solution of Eq.\,(\ref{gpessss}) 
has to vanish, as we have also checked numerically. This is simply due to the fact that if one 
goes to the rotating frame of the solitary wave, there is no time dependence and therefore 
$\langle \hat{\ell} \rangle$ has to be time independent.
\bibitem{comm} The case of a static wave corresponds to a ``grey" solitary wave, with a nonzero 
density minimum. It is only in the limit of an infinite ring, i.e., in the limit of $R \to \infty$ 
and $N \to \infty$, with $N/R$ finite, where the dark solitary wave is also static \cite{aaa}.
\bibitem{miner} In other superfluid systems, like nuclei \cite{Ben} and Bose-Einstein condensed 
atoms confined in harmonic potentials \cite{perez, Stringari} the moment of inertia vanishes in 
the presence of axial symmetry and it takes a nonzero value when the axial symmetry is broken.
\end{thebibliography}
\end{document}